\documentclass[aps,pra,twocolumn,superscriptaddress,showpacs,showkeys]{revtex4}
\usepackage[dvips]{graphicx}
\usepackage{longtable}
\usepackage{bm}
\DeclareBoldMathCommand{\bfmu}{\mu} 
\def\mum{\rm\mu m}                  
\def\muA{\rm\mu A}                  
\def\muK{\rm\mu K}                  
\def\ie{{\it i.e.,\/}}              
\def\eg{{\it e.g.,\/}}              
\def\etc{{\it etc.\/}}              
\def\GP{Gross-Pitaevskii}           
\def\CP{Casimir-Polder}             

\begin{document}

\title{Trapping cold atoms using surface-grown carbon nanotubes}

\author{ P. G. Petrov }
	\altaffiliation{Present address: Centre for Cold Matter, Blackett Laboratory, Imperial College London, United Kingdom}
	\affiliation{Department of Physics, Ben-Gurion University, Be'er Sheva 84105, Israel}
\author{ S. Machluf }
	\affiliation{Department of Physics, Ben-Gurion University, Be'er Sheva 84105, Israel}
\author{ S. Younis }
	\affiliation{Department of Physics, Ben-Gurion University, Be'er Sheva 84105, Israel}
	\affiliation{The Weiss Family Laboratory for Nano-Scale Systems, Ben-Gurion University, Be'er Sheva 84105, Israel}
\author{ R. Macaluso }
	\affiliation{Department of Physics, Ben-Gurion University, Be'er Sheva 84105, Israel}
	\affiliation{The Weiss Family Laboratory for Nano-Scale Systems, Ben-Gurion University, Be'er Sheva 84105, Israel}
\author{ T. David }
	\affiliation{Department of Physics, Ben-Gurion University, Be'er Sheva 84105, Israel}
\author{ B. Hadad }
	\affiliation{The Weiss Family Laboratory for Nano-Scale Systems, Ben-Gurion University, Be'er Sheva 84105, Israel}
\author{ Y. Japha }
	\affiliation{Department of Physics, Ben-Gurion University, Be'er Sheva 84105, Israel}
\author{ M. Keil }
	\thanks{Corresponding author}
	\email{keil@bgu.ac.il}
		\affiliation{Department of Physics, Ben-Gurion University, Be'er Sheva 84105, Israel}
\author{ E. Joselevich }
	\affiliation{Department of Materials and Interfaces, Weizmann Institute of Science, Rehovot 76100, Israel}
\author{ R. Folman }
	\affiliation{Department of Physics, Ben-Gurion University, Be'er Sheva 84105, Israel}

\begin{abstract}
We present a feasibility study for loading cold atomic clouds into magnetic traps created by single-wall carbon nanotubes grown directly onto dielectric surfaces. We show that atoms may be captured for experimentally sustainable nanotube currents, generating trapped clouds whose densities and lifetimes are sufficient to enable detection by simple imaging methods. This opens the way for a novel type of conductor to be used in atomchips, enabling atom trapping at sub-micron distances, with implications for both fundamental studies and for technological applications.
\end{abstract}

\date{\today}


\maketitle


\section{Introduction\label{sec:introduction}}

In the last decade the coherent manipulation of cold atoms has been reduced to the micrometer scale by realizing magnetic microtraps on dielectric substrates using standard microelectronic fabrication techniques.  These platforms, called atomchips~\cite{folman1,reichel1,muller,dekker}, enable the engineering of complicated potentials for manipulating atomic quantum states, including beamspitters, interferometers, lattices \etc~\cite{folman2,reichel,fortagh}. Bringing the atoms close to the atomchip surface, near the sources of these potentials, enables tight traps with low power consumption, and may enable a new tool for fundamental studies as well as numerous applications such as clocks, sensors and quantum information processing. However, as the atoms approach the dielectric or metallic surface, they are perturbed by atom-surface interactions and by temporal and spatial magnetic field fluctuations. On the one hand, this enables surface microscopy studies using ultra-cold atoms~\cite{wildermuth,aigner,japha} and studies of dispersion forces, including the \CP\ interaction~\cite{casimir,cornell,vuletic,sukenik,dalvit}, but on the other hand this destroys atomic coherence and introduces heating, trap loss and potential corrugation~\cite{henkel,jones,fortagh1}. 

There have been many suggestions for ways to overcome these limiting processes and experiments to quantify their success~\cite{rekdal,scheel,trebbia}. A recent proposal suggests that using electrically anisotropic materials can help reduce decoherence due to the nearby surface~\cite{tal}.  Another proposal is to employ metallic alloys at cryogenic temperatures to improve the lifetime of the trapped atomic samples~\cite{valery1}.  Utilizing superconducting wires in atomchips~\cite{valery2} may suppress  some of the hindering effects noted above. Such wires have recently been used to achieve Bose-Einstein condensation~\cite{roux}, to trap ultra-cold atoms using persistent currents~\cite{mukai}, and to study the Meissner state~\cite{cano}.

\begin{figure}
  \includegraphics[width=0.35\textwidth]{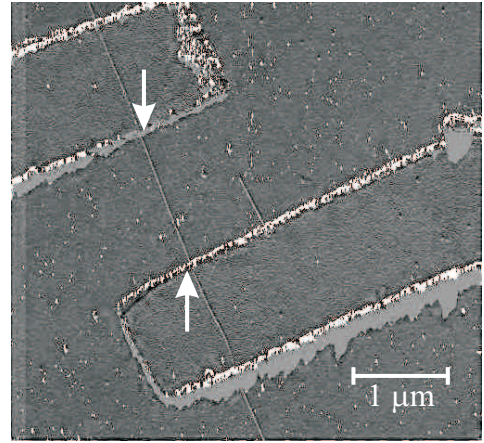}
  \caption{Atomic force microscope image of a~CNT fabricated and contacted for use as a ``Z''-shaped wire trap for atomchip experiments. This sample was fabricated as part of our fabrication feasibility study~\cite{macaluso}. It includes a straight~$1\,\mum-$long~CNT, grown on an~SiO$_2$ layer and electrically connected with~Pd wires ($1\,\mum$~wide $\times\ 25\rm\,nm$~thick). We have verified that the currents used for the simulations described in this work are sustainable for the repeated cycling required by atomchip experiments.
  \label{fig:CNT_el}}
\end{figure}

It has also been suggested that suspended carbon nanotubes~(CNTs) could prove to be advantageous for trapping ultra-cold atoms~\cite{peano,fermani}. In particular, these authors investigated suspended~CNTs in order to examine the feasibility of trapping cold atoms much closer than~$1\,\mum$ from the~CNT; since nearby surfaces were excluded, only the atom-nanotube \CP\ force needed consideration. For a~CNT lying directly on a dielectric surface and not suspended above it, the \CP\ force becomes much stronger because of the much larger surface area of the nearby substrate. This attractive force can destroy the atom trap by inducing tunneling through the magnetic potential barrier created by the current flowing through the~CNT. Nevertheless, since~CNTs grown on surfaces can carry higher currents and utilize simpler fabrication procedures, it is exactly this system for which we study the feasibility of trapping cold atoms at sub-micron distances. We also use accurate \GP\ calculations for atomic densities in order to fully estimate their trapping capabilities. 

The motivation to trap cold atoms close to a~CNT ranges from improving atom optics technology to fundamental studies of~CNTs.

The first motivation relates to the goal of creating a trap hundreds of nanometers from the surface {\it en route} to a real solid-state device with long coherence times~\cite{folman2,reichel,fortagh}. Such traps would enable high resolution manipulations of the external degrees of freedom (\eg\ creation of controllable tunneling barriers), high gradient traps with low power consumption, and small inter-trap distances for atom-atom entanglement (\eg\ for creating 2-qubit gates).  Achieving this goal may be hindered by the \CP\ force that attracts atoms to the surface at short distances, and by thermal noise that causes the atoms to undergo spin flips, heating, and decoherence, all of which may be reduced by using~CNTs~\cite{peano,fermani}.  CNTs may also offer less magnetic potential corrugation due to their ballistic transport of electrons~\cite{javey1}. In addition, when integrated with high-Q photonics, CNTs offer sharp absorption peaks (relative to metals) allowing resonator modes to be positioned in their vicinity.  Another possible advantage lies in the ability of~CNTs to form mechanical oscillators, thus providing coupling between a cold atom and a macroscopic device~\cite{treutlein}.  Finally, in the longer term, CNTs open the door to combining the field of atomchips with molecular electronics and self-assembled circuits.

The second motivation relates to the atoms' sensitivity to current corrugation~\cite{aigner,japha} and electron thermally induced currents~\cite{henkel,valery1}. This may enable experimental insight into the rich phenomena of electron transport in~CNTs and quantum effects such as orbiting electrons~\cite{refael} and spin-orbit coupling~\cite{kuemmeth}.  For example, measurements of the trap lifetime (limited by thermal~CNT noise-induced spin flips) will readily verify whether the standard theory for thermal magnetic noise in metals~\cite{henkel,folman2} is applicable to~CNTs.  In addition, the atoms may serve as a probe for forces such as \CP~\cite{cornell} and may therefore probe the forces induced by the~CNT.

In this paper we present a realistic scheme, based entirely on procedures available in atomchip fabrication facilities, for building and testing a single-wall~CNT atom trap. This will hopefully provide a first step in a practical effort eventually leading to much more complex geometries, including multi-wall CNT contacts with no need for metallic contacts in the vicinity of the atom trap~\cite{gao}, arrays of CNTs~\cite{ismach,abrams,geblinger}, and hopefully even deterministic growth or positioning of CNTs on atomchips~\cite{ismach,huang}. This feasibility study utilizes single-wall~CNTs grown directly onto a dielectric substrate and contacted by metal leads. We find that this configuration demands simpler fabrication, and as noted earlier, it will enable more complex geometries and larger currents than those possible for suspended nanotubes~\cite{pop}.  For example, to the best of our knowledge, the maximum current thus far achieved for suspended~CNTs of length~$L$ is~$\sim(10/L)\,\muA$~\cite{pop}, compared to~$20\,\muA$ ($L\geq3\,\mum$) and~$45\,\muA$ ($L\leq1.5\,\mum$) for surface-grown CNTs~\cite{yao,maune}. To verify simplicity in fabrication, we have also conducted a fabrication feasibility study, the detailed results of which will be made available elsewhere~\cite{macaluso}. In Fig.~\ref{fig:CNT_el} we present such a trap which we have fabricated, and on which the following simulations are based. Indeed, we have also verified with our fabricated sample that the currents used in these simulations are sustainable.

The paper is organized as follows.  In Sec.~\ref{sec:atomchip_design} we present details of the simple but practical design that we utilize for simulating a~CNT atomchip trap.  In Sec.~\ref{sec:ground_state} we calculate the ground state of the trapped atomic cloud using the \GP\ equation and a potential that includes the \CP\ interaction between the ultra-cold atoms and the surface of the atomchip substrate.  We estimate the trap lifetime in Sec.~\ref{sec:lifetime} by analyzing losses due to Majorana spin-flips, tunneling to the surface, thermal spin-flips, and atom heating.  In Sec.~\ref{sec:loading} we discuss loading the ultra-cold atomic cloud into the~CNT trap, followed by its release and detection using standard absorption imaging methods.  We summarize our results in Sec.~\ref{sec:conclusion}.


\section{Atomchip design\label{sec:atomchip_design}}
\begin{figure}
  \includegraphics[width=0.45\textwidth]{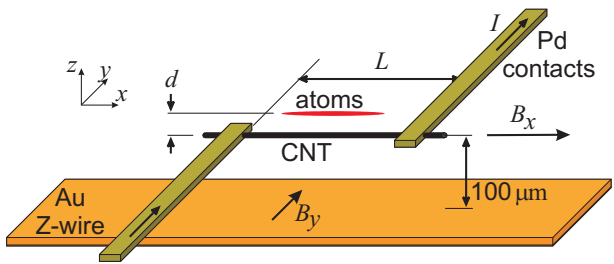}
  \caption{(Color online) Schematic representation of the~two-layer CNT atomchip, chosen as the basis for our feasibility study due to its simplicity. The nanotube is grown by chemical vapor deposition directly onto an upper~$\rm100\,\mum$-thick Si/SiO$_2$ substrate (the substrate is not shown in the schematic for clarity). This substrate is glued to a lower atomchip on which a gold~``Z''-wire has already been fabricated in order to facilitate loading the CNT trap. For clarity, only the central portion of this~``Z'' wire is shown. The lower chip contains additional wires for generating the necessary bias fields. The trap center is located at a distance~$d$ above the~CNT. Current flows in the direction designated by~$I$ through~Pd contacts that are separated by a distance~$L$. The~$+z$-axis is oriented in the direction of gravity.
  \label{fig:atomchip_schematic}}
\end{figure}

Throughout this paper we consider~$\rm^{87}Rb$ atoms trapped in the $\left|F=2,m_F=2\right\rangle$ ground state above a straight carbon nanotube that is grown on a~SiO$_2$-coated Si surface using standard chemical vapor deposition. The typical fabrication process we use produces straight~CNTs varying in length from~0.8 to~$>20\,\mum$~\cite{macaluso}; most of the calculations conducted in this study assume that the nanotube is~$5\,\mum$ long. The~CNT is electrically contacted with parallel~Pd leads~\cite{macaluso}, forming the atomchip trapping wire shown schematically in Fig.~\ref{fig:atomchip_schematic}. The resulting~``Z''-shaped wire (with~90$^\circ$ angles) used for this study is perhaps the simplest form for magnetic trapping~\cite{folman2}. We usually assume a~CNT current of~$20\,\muA$; noting that currents~$>40\,\muA$ have been achieved experimentally~\cite{maune}, we perform some calculations at~$35\,\muA$ in order to realistically characterize the nanotube trap sensitivity to current.

As with larger magnetic ``Z''-traps, we apply a small bias field~($B_y$) in the direction perpendicular to the nanotube axis, whose strength controls the position~$d$ of the trap minimum above the nanotube. A separate bias field~($B_x$) directed parallel to the nanotube axis allows adjustment of the magnetic field at the trap minimum (the Ioffe-Pritchard field~$B_0$), which is important for limiting the Majorana spin-flip rate (Sec.~\ref{subsec:majorana}).

The bias fields can be achieved by wires positioned on the lower chip of a double-layered chip design~\cite{gunther}, an example of which is shown in Fig.~\ref{fig:atomchip_schematic}. 
The vertical separation is necessitated by the fact that the magnetic field component parallel to the atomchip surface vanishes at the height of the wires producing it, so the~$B_x$ and~$B_y$ bias fields that are required very close to the~CNT cannot be created by wires placed in the same vertical plane. We use the double-layer configuration to simulate realistic bias fields. These fields are nearly homogeneous close to the CNT trap minimum but they are not as uniform as fields provided by external coils; we do not use the latter however, because the nanotube magnetic trap requires bias fields about three orders of magnitude smaller than typical bias fields. For such a high degree of spatial and amplitude control, one would like the source to be nearby. As presented in Fig.~\ref{fig:atomchip_schematic}, the lower chip also includes the loading wire made with standard gold patterning techniques, while the~CNT is visible on the upper chip (not shown). The double-layer design also allows implementation of two very different fabrication processes and accurate alignment of the gold loading wire directly beneath the~CNT~\cite{macaluso} for optimizing the loading procedure (Sec.~\ref{sec:loading}). Our simulations assume a~$100\,\mum$ vertical distance between the loading and~CNT ``Z'' wires, based on commercially available~Si/SiO$_2$ substrates on which we have already grown suitable CNTs~\cite{macaluso}.

\begin{figure*}
  \includegraphics[width=1.0\textwidth]{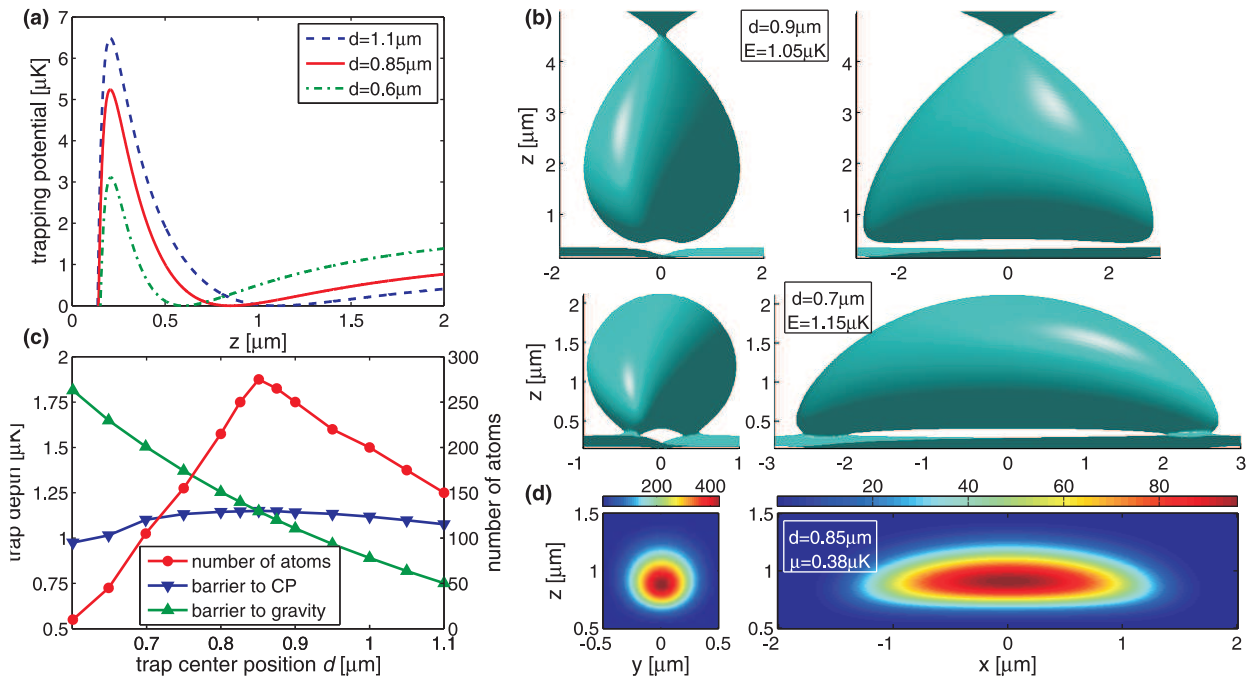}
  \caption{(Color online) Atomic trapping potentials and optical densities for CNT traps calculated using a current of~$20\,\muA$ through a~$5\,\mum$-long nanotube.
  (a)~Trapping potential directly above the nanotube~($x=y=0$) as a function of the height~$z$ for different positions~$d$ of the trap center. The trap center is controlled using the bias field~$B_y$.
  (b)~Isopotential surfaces viewed along the~$x-$ and~$y-$axes at an energy of~$1.05\,\muK$ for a trap centered at~$d=0.9\,\mum$ and~$1.15\,\muK$ for~$d=0.7\,\mum$. The openings show where atoms can escape due to gravity or to the surface, respectively.
  (c)~Trap depth (triangles) and number of trapped atoms (circles) as a function of the trap center position~$d$. Inverted triangles show the energy above which atoms can reach the surface due to the \CP\ force; upright triangles show the energy above which atoms can drop out of the trap due to gravity.
  (d)~{\it In~situ} optical density maps of the trapped atoms for~$d=0.85\,\mum$, as viewed parallel to the nanotube and perpendicular to it, calculated using the \GP\ equation for a trap containing~275 interacting ground state atoms; the chemical potential~$\mu$ is~$1/3$ of the trap depth at this height.
  \label{fig:at_pot}}
\end{figure*}

\section{Ground state of the CNT trap\label{sec:ground_state}}
In this section we present a numerical analysis of various~CNT trap potentials, based on solutions to the \GP\ equation. The chemical potential and atom density are calculated for ground-state atoms and the latter is used to estimate the expected optical density of the atomic cloud.

\subsection{Atomic trapping potential\label{subsec:at_pot}}
Nanotube magnetic guides are expected to enable very strong confinement in the transverse direction when atoms are brought nearby~\cite{peano}. At the same time however, the \CP\ short-range attractive force~\cite{casimir} limits how close the atoms may approach the~CNT without being lost to it by tunneling through the magnetic potential barrier created by the current through the~CNT. The trap gradient increases when atoms are brought closer to the~CNT, but the barrier height is reduced. In our case, where we have chosen the simplest configuration in  which the~CNT is not suspended, atoms approaching the~CNT also feel the \CP\ force due to the substrate surface. Throughout this paper, we will consider only this atom-surface \CP\ force since it is expected to be much larger than the atom-CNT force.

For a ground state atom at a distance~$z$ from a planar dielectric surface of static permittivity~$\epsilon$, the \CP\ potential can be written in the form~\cite{antezza2}
\begin{equation}\label{eq:cp_pot}
    U_{\rm CP}=-\frac{C_4}{z^4},~~{\rm where}~~C_4=\frac{3}{8}~\frac{\hbar
    c\alpha_0}{\pi}~\frac{\epsilon-1}{\epsilon+1}~\phi(\epsilon),
\end{equation}
and~$\alpha_0=47.3\times10^{-24}\,\rm{cm^3}=5.25\times10^{-39}\,\rm{F\cdot m^2}$ is the ground state polarizability of the~$\rm^{87}Rb$ atom~\cite{antezza2}. The calculation of the dimensionless function~$\phi(\epsilon)$ for a single dielectric substrate is described in detail in~\cite{yan}. In our case however, the substrate is a dielectric bilayer consisting of a~$100\,\mum$-thick Si~wafer coated by a~$\rm200\,nm$-thick layer of~SiO$_2$ required to avoid shorts and to improve~CNT growth. Under these circumstances, the~CP potential is dominated by~Si for large~$z$ ($>1\,\mum$), while for small~$z$ the potential is dominated by the thin~SiO$_2$ layer. For the intermediate values of~$z$ needed here, a more general description appropriate for multilayered dielectrics is required~\cite{zhou}; this can be approximated by replacing~$\phi(\epsilon)$ with a generalized function of both dielectrics and the thickness of the upper layer based on the formalism of Schwinger~\cite{schwinger,milton}
\begin{equation}\label{eq:cp_milton}
   \frac{\epsilon-1}{\epsilon+1}~\phi(\epsilon)~\longrightarrow~F(\epsilon_1,\epsilon_2,t,z),
\end{equation}
where~$\epsilon_1=4$ and~$\epsilon_2=12$ for~SiO$_2$ and~Si respectively, and~$t=200\rm\,nm$ is the thickness of our upper~SiO$_2$ layer. Milton~\cite{milton-pc} has calculated the function~$F$ for these parameters, and we find that it can conveniently be fit by the expression~$F(\epsilon_1,\epsilon_2,t,z)=ae^{-b/z}+c$ where~$a=0.223$, $b=0.822\,\mum$, and~$c=0.463$. At a trapping height of~$z=0.85\,\mum$ (measured from the top surface), this yields~$F=0.55$~\cite{milton-pc}, about~18\% larger than for pure~SiO$_2$, so thicker layers of~SiO$_2$ would not reduce the \CP\ force significantly (a thicker layer of~SiO$_2$ would however reduce the heat conductivity significantly).  Other substrates commonly used for~CNT growth, such as sapphire~\cite{ismach}, have much higher dielectric constants ($\epsilon=9.3-11.5$) so their use would also not reduce the \CP\ force.

The \CP\ interaction is significant only for small atom-surface distances, modifying the potential created by the interaction of the atomic magnetic dipole moment~$\bfmu$ with the magnetic field~$\mathbf{B}(x,y,z)$. The total potential is then the sum of the magnetic potential~$U_{\rm mag}(x,y,z)=-\bfmu\cdot\mathbf{B}(x,y,z)$, the \CP\ potential~$U_{\rm CP}$, and the gravitational potential~$U_{\rm grav}=+mgz$ (the atomchip lies above the trap, in the~$-z$ direction as defined in Fig.~\ref{fig:atomchip_schematic}, in order to allow free-fall upon release for detection):
\begin{equation}\label{eq:tot_pot}
    U(x,y,z)=U_{\rm mag}(x,y,z)+U_{\rm CP}(z)+U_{\rm grav}(z).
\end{equation}

Examples of the atomic trapping potential are shown in Fig.~\ref{fig:at_pot}(a) for various trap center positions~$d$. It is evident that optimizing~$d$ to obtain the deepest trap requires balancing the potential barrier at low heights~$z$ with the barrier at large~$z$ that is provided by the bias field~$B_y$. To find this balance accurately, we must also consider the trap shape, two examples of which are shown by the isopotential surfaces in Fig.~\ref{fig:at_pot}(b). Here we see more clearly the contrast between the high gradients near the atomchip surface and the weaker gradients further away, causing the broad bulge for large~$z$. Also apparent is a pronounced bending of the potential towards the surface at both ends of the trap.

The isopotential surfaces in Fig.~\ref{fig:at_pot}(b) show two possible escape routes, away from the surface due to gravity directly above the center of the trap (for~$d=0.9\,\mum)$, or towards the surface due to the Casimir-Polder interaction at the ends of the trap (for~$d=0.7\,\mum$). These escape routes are most efficiently blocked by two balanced barriers at a trap height of~$d=0.85\,\mum$, where the two curves of Fig.~\ref{fig:at_pot}(c) intersect. This height then corresponds to the deepest possible trap of~$1.15\,\muK$ for the present configuration, which in turn optimizes the number of atoms that can be held in the trap, as discussed in the following section.

\subsection{Atomic and optical densities\label{subsec:OD}}

The ground state~$\psi(\mathbf{r})$ of~$N$ interacting bosons in an external potential is given by the \GP\ equation~\cite{dalfovo}:
\begin{equation}\label{eq:GP}
\left[-\frac{\hbar^2}{2m}\nabla^2+V(\mathbf{r})+g|\psi(\mathbf{r})|^2\right]\psi(\mathbf{r})=\mu\psi(\mathbf{r}),
\end{equation}
where~$m$ is the atomic mass,~$V(\mathbf{r})$ is the external potential,~$\mu$ the chemical potential, and~$g=4\pi\hbar^2a/m$ is the coupling constant, with~$a$ being the {\it s}-wave scattering length ($a=\rm5.4\,nm$ for~$\rm^{87}Rb$).

For systems with a ``large'' number of interacting atoms, one often uses the Thomas-Fermi approximation~\cite{dalfovo}. As an example, for~275 atoms in a~$5\,\mum$-long nanotube trap, this approximation underestimates the chemical potential by~$\sim15\%$ and the corresponding wavefunction is~$\sim10\%$ more confined in the transverse direction (at~10\% of the peak probability) than the \GP\ wavefunction. We therefore solve the full \GP\ equation for the atomic density calculations throughout this study. For a given number of atoms in the~CNT trap, these solutions yield a ground state energy, which can be expressed as a fraction of the trap depth. Conversely, fixing the ratio of the trap depth to the ground state energy determines the number of atoms~$N$ that may be held by any particular~CNT trap. We fix this ratio at~$3.0$ for all the \GP\ calculations discussed in this paper in order to ensure adequate trapping.

The optical density of the atomic cloud is an essential parameter for considering detection by standard imaging techniques, as will be discussed in Sec.~\ref{sec:loading} below. We calculate the ground state density distribution of~$N$ interacting atoms for a nanotube trap of given length and current and then convert this to the optical density for resonant absorption detection by integrating along the~$x$ or~$y$ directions and multiplying by the absorption cross-section $\sigma=2.907\times10^{-9}\,\rm cm^{-2}$, as presented in Fig.~\ref{fig:at_pot}(d).

Trapping atomic clouds above~CNTs can be a challenging task due to the limited current that can be sustained through the nanotube, their shortness, and the \CP\ attractive force to the surface (see also Sec.~\ref{sec:lifetime}). It is natural to propose that increasing the trap volume would enable more atoms to be collected. Using the \GP\ equation and the above calculations of optical density, we therefore investigate properties of~CNT traps using longer nanotubes and higher currents, while remaining within practical limits provided by our fabrication feasibility study~\cite{macaluso}.

\begin{table*}
  \caption{Atomic cloud parameters calculated as a function of the nanotube length~$L$ and the current~$I_{\rm CNT}$. In each case the trap center position~$d$ is chosen to ensure the deepest trap, and the Ioffe-Pritchard field~$B_0$ is chosen so that the Majorana spin-flip rate is~$\rm1\,Hz$ (Sec.~\ref{subsec:majorana}).  
  \label{tbl:length_dependence}}
  \begin{ruledtabular}
  \begin{tabular}{c c c c c c c c c c c c}
$L$       &$I_{\rm CNT}$  &$d$       &$B_0$  &$N$ \footnotemark[1] 
                                                   &$\mu$  &Trap depth  &$\Delta x$ \footnotemark[2] 
                                                                                    &$\Delta z$ \footnotemark[2]
&OD$_y$ \footnotemark[3]                                                  
          &$\nu_{r}$ \footnotemark[4]          
                          &$\nu_{ax}$ \footnotemark[4]                                           \\
$(\mum$)  &$(\muA)$       &$(\mum)$  &(mG)   &     &(kHz)  &$(\muK)$    &($\mum)$   &($\mum)$        &         &(kHz)          &(kHz)                                                                \\
\hline
1         &35             &0.65      &67.7   &15   &11.2   &1.66        &0.7        &0.4           &40       &8.8            &3.0                                                                  \\
3         &35             &0.80      &50.2   &340  &16.0   &2.31        &1.8        &0.60             &205      &6.7            &1.35                                                                 \\
5         &35             &0.90      &42.1   &1000 &16.4   &2.37        &3.3        &0.75         &270      &5.7            &0.73                                                                 \\\\
5         &20             &0.85      &30.5   &275  &8.0    &1.15        &2.8        &0.65             &100      &4.2            &0.54                                                                 \\
10        &20             &0.85      &30.1   &820  &8.2    &1.18        &6.9        &0.69             &110      &4.2            &0.18                                                                 \\
15        &20             &0.85      &30.0   &1500 &8.0    &1.15        &11.4       &0.75             &105      &4.1            &0.09                                                                 \\
20        &20             &0.85      &29.9   &2000 &7.8    &1.12        &16.3       &0.73             &95       &4.1            &0.05                                                                 \\
  \end{tabular}
  \end{ruledtabular}
\footnotetext[1]{ Number of atoms, adjusted for each~$L$ and $I_{\rm CNT}$ so that the chemical potential~$\mu$ is~1/3 of the trap depth.}
\footnotetext[2]{ Distance between points at which the optical density drops to~$e^{-2}$ of its maximum value.} 
\footnotetext[3]{ Peak optical densities are given for imaging along the~$y$-axis (Fig.~\ref{fig:atomchip_schematic}).} 
\footnotetext[4]{ Radial and axial trap frequencies.}
\end{table*}

In Table~\ref{tbl:length_dependence} we present calculated atom numbers and trap characteristics for three ``short'' ($1-5\,\mum$) nanotube traps operating at~$35\,\muA$, and for four ``long'' ($5-20\,\mum$) traps operating at~$20\,\muA$. The expected optical densities are well within those typically observed in~BEC experiments. As can be seen, the radial trapping frequency~$\nu_r$ and the optical density~OD$_y$ depend on the~CNT current but not on its length (except for the shortest nanotubes). The axial trapping frequency~$\nu_{ax}$ drops in proportion to the trap length since the contacts that provide the longitudinal confinement are further apart. For both currents considered, the atom number~$N$ grows with increasing nanotube length~$L$, reaching~2000 atoms for a~$20\,\mum$ long nanotube operating at~$20\,\muA$. Increasing the current to~$35\,\muA$ results in tighter traps, as shown by the higher radial frequencies~$\nu_r$, and sharply increases the atom number and the corresponding optical density (\eg\ almost a four-fold increase for the~$5\,\mum$ long nanotube). This should encourage experimental improvements in the maximum current of~CNTs, \eg\ by decreasing contact resistance or by improving phonon-mediated heat transfer to the substrate~\cite{kooi}.

\section{Lifetime of the CNT trap\label{sec:lifetime}}
In typical atomchip experiments atoms are trapped close to a metallic surface. Random thermal magnetic field fluctuations caused by Johnson noise within nearby conductors introduce heating, trap loss, and decoherence, even if the conductors are not carrying current. Technical imperfections, such as unstable current supplies, introduce further heating. Tunneling through the magnetic potential barrier, caused by the atom-surface \CP\ interaction~(Fig.~\ref{fig:at_pot}), also limits the trap lifetime. This situation is very different from that encountered with  suspended~CNTs~\cite{fermani}, for which the much weaker atom-CNT \CP\ force causes much slower tunneling loss. Independent of the surface are trap losses caused by Majorana spin flips. Other limits to the trap lifetime, \eg\ losses due to background gas collisions, do not contribute significantly since they are typically longer under realistic experimental conditions.

\subsection{Majorana spin-flip rate\label{subsec:majorana}}
Cold atoms in a low-field seeking state that are trapped near a vanishing magnetic field can undergo a spin-flip transition to a high-field seeking state that is untrapped (Majorana spin-flips). Applying a small offset (Ioffe-Pritchard) field~$B_0$ at the trap center reduces the spin-flip transition rate as given by the approximate formula~\cite{sukumar}:
\begin{equation}\label{eq:Majorana_spin_flip_rate}
    \Gamma_{\rm M}=\frac{\pi\omega_r}{2}~
    \exp\left(-\frac{2|\bfmu||\mathbf{B_0}|
    +\hbar\omega_r}{2\hbar\omega_r}\right),
\end{equation}
where~$\omega_r$ is the trap radial frequency. Equation~(\ref{eq:Majorana_spin_flip_rate}) is valid when the Larmor frequency $\omega_{\rm L}=|\bfmu||\mathbf{B_0}|/\hbar\gg\omega_r$, requiring that~$B_0\gg\rm5\,mG$ for radial frequencies typical of the nanotube traps we are considering. Under these conditions, we choose a Ioffe-Pritchard field~$B_0$ that yields a Majorana spin-flip loss rate of~$\rm1\,Hz$ for each of the nanotube traps shown in Fig.~\ref{fig:at_pot}(b-d) and characterized in Table~\ref{tbl:length_dependence}.

\subsection{Tunneling rate\label{subsec:tunneling}}

As seen from Fig.~\ref{fig:at_pot}(a-c), atoms can tunnel through the finite barrier to the atomchip surface. The single-atom tunneling rate in one dimension is given by the following expression~\cite{CT_QM}:
\begin{equation}\label{eq:tunnel_rate}
    \Gamma_{{\rm tunn}}=\nu_{r}
    ~\exp\left(-2\int_{z_{1}}^{z_{2}}dz
    \sqrt{\frac{2m}{\hbar^2}\left[U(0,0,z)-\mu\right]}
    \right),
\end{equation}
where the integration is between the classical turning points~$z_1$ and~$z_2$. As discussed in Sec.~\ref{subsec:at_pot}, we assume that the \CP\ attraction to the~CNT itself is much weaker than to the dielectric surface, and hence we calculate the tunneling rate accounting only for the latter. A comparison of the Majorana spin flip rate to the tunneling rate is presented in Fig.~\ref{fig:lifetime}. At the closest distance of~$d=0.6\,\mum$, we calculate $\Gamma_{\rm tunn}=\rm0.06\,Hz\ll\Gamma_{\rm M}$, fixed at~$\rm1\,Hz$ as above. Since the tunneling rate drops rapidly as~$d$ is increased, we do not expect this to be a major loss mechanism for any of the traps considered in this paper.

\subsection{Thermal noise-induced spin flips\label{subsec:thermal}}

The coupling of cold atoms to thermal near-field radiation, arising from the random motion of electrons within a nearby metallic surface (Johnson noise), leads to trap loss due to spin flips (as well as to heating and decoherence). This loss becomes larger as the atom-surface distance is decreased, and occurs even without an applied current in the conductor. In typical atomchip experiments the spin-flip loss rate due to such intrinsic noise dominates that due to technical noise~\cite{folman2}; the latter leads to trap loss mainly through heating and is discussed in the next section. Assuming that the theory of noise from metals is applicable to~CNTs, the thermal spin flip rate can be written as~\cite{tal,henkel}
\begin{equation}\label{eq:spinfliprate}
\Gamma_{\rm th} =
\mu_{\rm B}^2g_F^2~
\frac{k_B T}{4\pi^2\hbar^2\;\epsilon_0^2\;c^4}
\sum_{l,m=\perp}
\left\langle i\left|F_l\right|f\right\rangle
\left\langle f\left|F_m\right|i\right\rangle~
Y_{lm},
\end{equation}
where we sum the contribution of all components of the noise perpendicular to the atomic magnetic moment, $F_l$~is the~$l^{\rm th}$ component of the spin operator, $T$~is the surface temperature, and $Y_{ij}={\rm tr}(X_{ij})\delta_{ij}-X_{ij}$ contains the geometrical integrals
\begin{equation}\label{eq:Xij}
X_{ij} =
\frac{1}{2}
\int_V {\rm d}\mathbf{x'}
\frac{\left(\mathbf{x}-\mathbf{x'}\right)_i
      \left(\mathbf{x}-\mathbf{x'}\right)_j}
     {\left|\mathbf{x}-\mathbf{x'}\right|^3
      \left|\mathbf{x}-\mathbf{x'}\right|^3}
\end{equation}
that sum up the contribution of local fluctuations arising from each point in the surface volume. This is calculated within the quasi-static approximation~\cite{tal,henkel2}, which is valid when~$d$ is smaller than the skin depth $\delta=\sqrt{2\rho/\mu_0\omega_{\rm L}}$ ($\rho$ is the resistivity and~$\mu_0$ is the permeability of free space). This condition is easily met here as~$d\lesssim1\,\mum$, much smaller than the skin depths of metals that are typically tens of~$\mum$ (the~Pd skin depth at the Larmor frequency is~$\delta\approx160\,\mum$). 

\begin{figure}
  \includegraphics[width=0.45\textwidth]{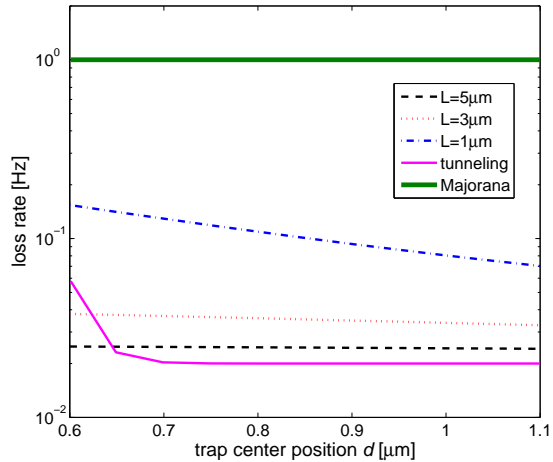}
  \caption{(Color online) Trap loss rates as a function of the distance~$d$ between the trap center and the surface due to Majorana spin flips, tunneling to the surface, and thermal noise-induced spin flips from the~Pd contacts, assuming a~$\rm50\,s$ lifetime due to collisions with background gases. The Majorana and tunneling loss rates are shown only for a~$5\,\mum$-long CNT; the thermal spin-flip rates are for~1, 3 and~$5\,\mum$-long CNTs. The thermal spin-flip contribution from the~CNT is negligible compared to the metal contacts because of its comparatively high resistivity. The chosen Majorana spin-flip rate limits the overall trap lifetime to~$\rm1\,s$ for all values of~$d$ considered in this feasibility study.
  \label{fig:lifetime}}
\end{figure}

The noise from the nanotube itself is expected to be very small~\cite{fermani}, resulting in loss rates below~$\rm0.1\,Hz$ even for~$d<1\,\mum$. The noise from the~Au loading wire, located~$100\,\mum$ beneath the~CNT trap, is also negligible (loss rate~$\rm<0.01\,Hz$). For our experimental design we thus expect loss rates due to thermal noise to be dominated by thermal radiation originating in the~Pd contacts, which are located~$<10\,\mum$ from the atoms. We therefore calculate the thermal noise-induced spin-flip loss rate according to Eq.~(\ref{eq:spinfliprate}) as a function of~$d$, for different lengths~$L$ of the nanotube, and display the results in Fig.~\ref{fig:lifetime}. We consider a cascade event of spin flips from $\left|F=2,m_F=2\right\rangle$, through~$\left|2,1\right\rangle$, to the untrapped state~$\left|2,0\right\rangle$, whereupon the atoms are assumed to be lost immediately. The total loss rate is an average over the entire trap volume, which we approximate by calculating the loss rate where the atom density is highest, namely at the trap center~\cite{fortagh}. The spin flip rate increases for shorter nanotubes, as expected when the distance to the metal contacts decreases. For the shortest nanotube~$L=1\,\mum$ and the closest distance of~$d=0.6\,\mum$ considered in this paper, we calculate $\Gamma_{\rm th}=\rm0.15\,Hz$; for~$L=3\,\mum$ this decreases to~$\Gamma_{\rm th}=\rm0.04\,Hz$ and for~$L=5\,\mum$ it drops by a further~50\%. We therefore do not expect thermal noise-induced spin flips to exceed those due to Majorana spin-flips, which we fix to be a constant as noted previously.

\subsection{Technical noise\label{subsec:technical}}

Another harmful mechanism to be considered is due to atom heating in the magnetic trap. Heating may lead to atoms acquiring enough kinetic energy to escape from the trap, especially because of the small trap depths being considered, or it may destroy coherence by causing the critical temperature for condensation to be exceeded. In typical atomchip traps the main source of heating is instability in the currents applied to the trapping micro-structure, \eg\ arising from imperfect power supplies (technical noise). Heating due to thermal noise from the surface is typically several orders of magnitude lower~\cite{folman2}. The single-atom heating rate~$\dot{T}$ arising from excitations of atoms from the ground vibrational state to the first excited state is given in convenient units by~\cite{Henkel1}
\begin{equation}
\dot{T} \approx 3\cdot10^9~
{\rm\frac{nK}{s}}
\frac{\hbar\omega}{k_{\rm B}}
\frac{m}{\rm amu}
\left(\frac{\omega}{2\pi\cdot100{\rm kHz}}\right)^3
\frac{I/{\rm A}}{(B_y/{\rm G})^2}
\frac{S_I(\omega)}{SN_I},
\label{Eq:HeatingRateTechNoise}
\end{equation}
where~$\omega$ is the trap frequency, $I$~is the current in the trapping micro-structure, and~$B_y$ is the homogeneous bias field in the direction perpendicular to the wire axis. 

The power spectrum~$S_I$(in units of~$\rm A^2/Hz$), characterizing the noise of the parallel component of the magnetic field, is related to the noise amplitude spectrum of the power supply by $S_v~[{\rm dBV/\sqrt{Hz}}] = 20\log_{10}(\sqrt{R^2[\Omega^2]\cdot S_I~[{\rm A^2/Hz}]})$, where $S_v=-140\rm\,dBV/\sqrt{Hz}$ is a typical value for commercial low-noise current sources and~$R$ is the load resistance in ohms. Only this component of~$S_I$ needs to be considered here because we are concerned only with magnetic fluctuations that do not flip the spin. Finally, $SN_I=0.16\rm nA^2/Hz\cdot(I/A)$ is the current shot-noise reference level. Because the resistance of the~CNT and its contacts is quite high (typically, $R\gtrsim\rm40\,k\Omega$), the level of~$S_I$ is actually shot-noise limited for the value of~$S_v$ chosen.

Taking the current, bias field, and trap frequency of our~$5\,\mum$-long nanotube trap operating at~$20\,\muA$ (Table~\ref{tbl:length_dependence}), we calculate a heating rate of $\dot{T}\approx\rm0.1\,nK/s$. Even for the shortest, highest-frequency traps considered, we calculate $\dot{T}\approx\rm0.5\,nK/s$ and we conclude that heating due to technical noise is not expected to limit the experimental lifetime. Note that these calculated heating rates are significantly smaller than previously measured rates~\cite{fortagh1,hansel}, as the technical noise here is limited to the shot-noise level because of the low current and high resistance of the~CNT.

\begin{figure}
  \includegraphics[width=0.45\textwidth]{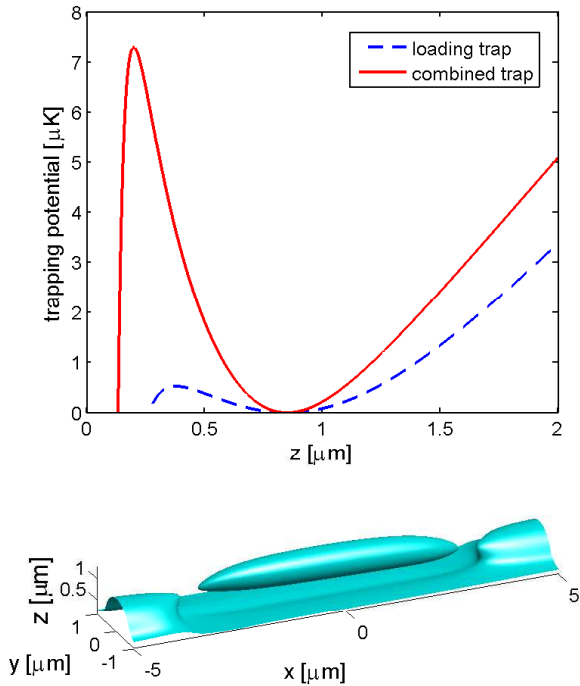}
  \caption{(Color online) 
  (a)~Atomic trapping potentials due to a current of~$\rm0.5\,A$ through the gold~``Z''-wire only (the ``loading'' trap), and due to the ``combined'' trap, which has in addition a~$20\,\muA$ current flowing through a~$5\,\mum$-long nanotube. Both trap minima are~$0.85\,\mum$ from the surface. The nanotube current creates a much higher potential barrier for the combined trap as well as generating tighter confinement. 
  (b)~Potential isosurface for the combined trap at an energy of~$1.3\,\muK$ using the above currents, demonstrating how the combined~CNT trap ``pinches off'' the atom density directly above the nanotube contacts and isolates it from the atomchip surface.
  \label{fig:pinching}}
\end{figure}

\begin{figure*}
  \includegraphics[width=1.0\textwidth]{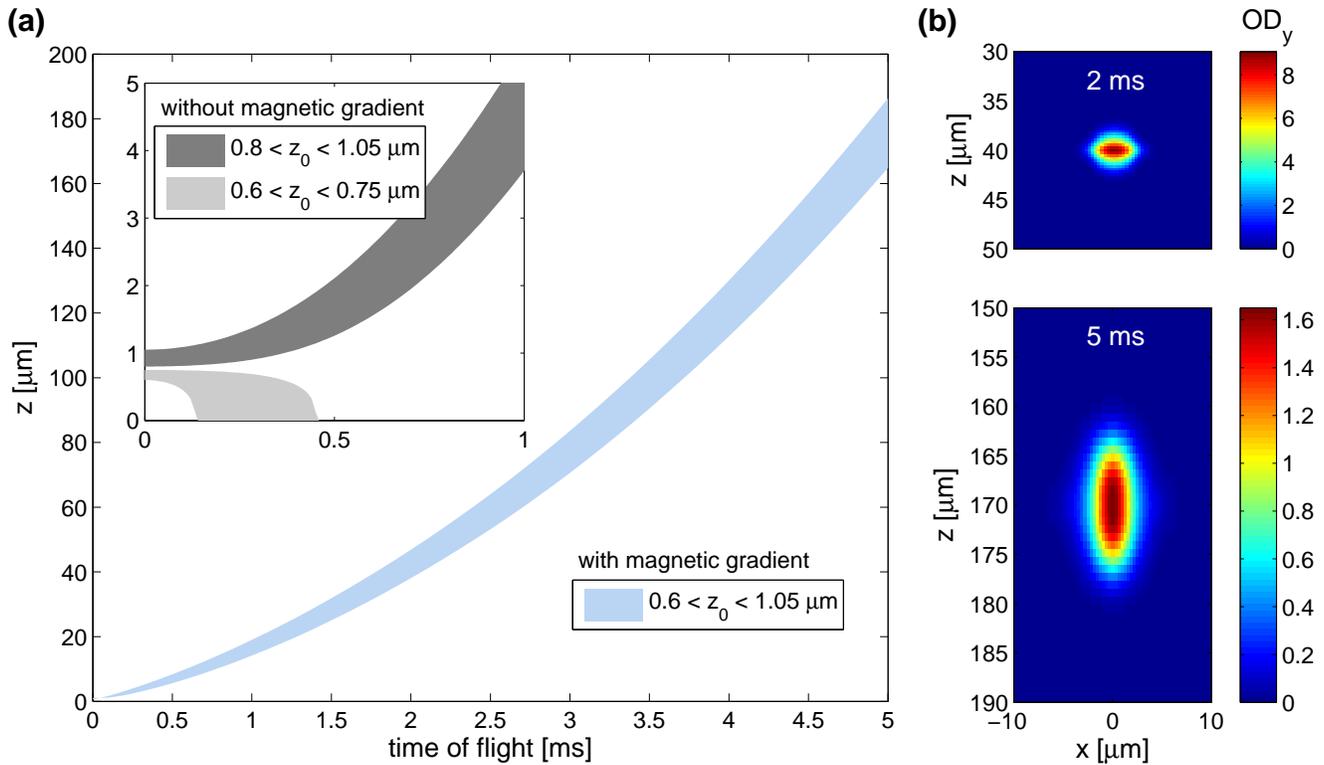}
  \caption{(Color online) 
  (a)~Position of atoms in the cloud as a function of time of flight under the influence of external potentials. Bands correspond to the range of initial starting points~$z_0$ within the cloud, whose center is located at~$d=0.85\,\mum$. The main graph shows the atomic positions under the influence of gravity, the \CP\ force, and a magnetic gradient. The inset shows the atomic positions, at short time-of-flight, under the influence of only gravity and the \CP\ force. Atoms at all points starting at~$z_0<z_{\rm eq}=0.76\,\mum$ will crash into the chip unless a magnetic gradient is applied. 
  (b)~Evolution of the atom cloud optical density after a time-of-flight of $\rm2\,ms$~(upper) and $\rm5\,ms$~(lower) under the influence of external potentials and atomic collisions. The cloud becomes resolvable after~$\rm2\,ms$ and remains detectable even after~$\rm5\,ms$ using standard imaging systems. See text for trap release details.
  \label{fig:imaging}}
\end{figure*}

\section{Loading and releasing\label{sec:loading}}

As part of this feasibility study, we also wish to analyze the loading of the~CNT trap. Transferring atoms from one magnetic trap to another requires optimization of bias fields and gradients such that the atom transfer is adiabatic, \ie\ the heating due to the transfer process is minimal. In experimental language this transfer is nearly adiabatic if the two traps are ``mode-matched''~\cite{ketterle}, \ie\ that the frequencies and position of the traps must be similar and the change of the magnetic field gradients should be such that atoms are not driven into collective oscillation.

Loading of the~CNT trap refers to the final transfer of atoms from a standard gold wire trap to the~CNT wire trap. Here, we make use of the simple gold ``Z''-shaped wire located~$100\,\mum$ directly below the~CNT on the lower layer of the atomchip (Fig.~\ref{fig:atomchip_schematic}). By applying a bias field perpendicular to the wire, a magnetic ``loading'' trap is created from which atoms can be transferred into the~CNT trap. At the beginning of the sequence, the loading trap is deep and relatively far from the atomchip surface, and current through the~CNT does not influence the trap. As we gradually move the loading trap closer to the nanotube by increasing the applied bias field, the magnetic field gradient increases, enabling ``mode-matched'' conditions. 

The trap gets shallower as the loading trap is brought closer to the atomchip surface, \eg\ from~350 to~$45\,\muK$ as the trap minimum position is decreased from~50 to~$10\,\mum$ from the nanotube~(150 to~$110\,\mum$ from the gold ``Z''-wire). Such a trap can still hold thermal atoms.  Below~$1\,\mum$ however, the barrier height for the loading trap is reduced to~$<1\,\muK$ by the \CP\ potential, as seen in Fig.~\ref{fig:pinching}(a). To overcome this problem we create a ``combined'' trap by running a current through the~CNT simultaneously.  At these close distances, a current of~$20\,\muA$ in the~CNT is sufficient to increase the barrier height of the combined potential to~$>7\,\muK$, thus providing a good shield against tunneling. Consequently, atoms that are immediately above the nanotube will be trapped as shown in Fig.~\ref{fig:pinching}(b) while the remainder, originally trapped by the (longer) gold ``Z''-wire, will be lost to the surface. The effect of the~CNT in the combined trap is to ``pinch off'' the atomic cloud directly above the nanotube contacts, thus isolating its central portion from surface depletion. Once this isolation is achieved, current in the gold ``Z''-wire can be turned off, leaving the atoms trapped by the magnetic field generated only by the nanotube.

Finally, after some trapping time, the atoms should be detectable. This could be accomplished using highly sensitive micro-cavity~\cite{trupke,purdy,steinmetz,lev}, micro-disc~\cite{rosenblit2,rosenblit1,rosenblit,aoki,barclay}, or fiber-based fluorescence~\cite{quinto-su,haase} techniques. In this feasibility study however, we focus on simple resonant absorption imaging using external optical elements in order to avoid further fabrication on the~CNT atomchip (although, if required, this can indeed be done). 

Given the typical spatial resolution of absorption imaging, we need to release the cloud from the~CNT trap and allow it to drop (due to gravity) and expand. The cloud should drop far enough that the imaging laser beam will not be diffracted from the surface ($\gtrsim50\,\mum$), and the cloud should expand sufficiently for it to be observable with modest resolution ($\sim2\,\mum$) without losing too much optical density ($\gtrsim0.1$). These conditions are satisfied within~$\sim2-5\rm\,ms$ time-of-flight, as shown in Fig.~\ref{fig:imaging}.

Trap release may be initiated by turning off all the magnetic fields and the~CNT current. However, if the magnetic force is turned off completely, the strong \CP\ force would overcome gravity for portions of the cloud that are released too close to the surface [Fig~\ref{fig:imaging}(a)]. While adiabatic de-loading of the magnetic trap is possible~\cite{vuletic}, we consider instead ejecting all the atoms by turning off the~$B_y$ and $B_x$ bias fields, thereby applying a magnetic gradient due to the current still passing through the~CNT. As both~$B_0$ and the~CNT field give rise to a Larmor frequency of only a few tens of~kHz, one should take care to close the fields sufficiently slowly to avoid spin flips to high-field seeking states. In early stages of the trap release, the bias fields are therefore turned off gradually while the magnetic field generated by the nanotube is still on. This creates a magnetic potential which is still trapping, but with reduced frequencies and which is moving progressively further from the atomchip, thereby repelling the atom cloud. Once the bias fields are turned off completely, the net force acting on the center of mass of the cloud is the sum of the attractive \CP\ force $\mathbf{F}_{\rm CP}=~-\nabla U_{\rm CP}$ dragging atoms toward the surface, the magnetic gradient force $\mathbf{F}_{\rm mag}$ repelling the atoms from the nanotube, and the gravitational force $\mathbf{F}_{\rm grav}=mg$, also directed away from the atomchip surface. The resulting equation for the center-of-mass motion is
\begin{eqnarray}\label{eq:z(t)-kick}
m\ddot{z}(t)+\frac{4C_4}{z(t)^5}-F_{\rm mag}-mg=0.
\end{eqnarray}

We solve the above differential equation classically for two cases, as shown in Fig.~\ref{fig:imaging}(a). We first consider atoms under the influence of gravity and the \CP\ force, but exclude the magnetic gradient. In this case, for atom-surface distances closer than the position of unstable equilibrium~$z_{\rm eq}=0.76\,\mum$, defined by the condition $F_{\rm CP}(z_{\rm eq})=F_{\rm grav}(z_{\rm eq})$, the net force exerted on the released atomic cloud is directed towards the chip surface. As a result, atoms released with~$z_0<z_{\rm eq}$ will collide with the atomchip unless we retain the magnetic gradient. We therefore also solve Eq.~(\ref{eq:z(t)-kick}) with all three potential terms. In order to evaluate how different parts of the cloud move, we assume a range of initial positions around the trap center at~$d=\rm0.85\,\mum$.

The inset of Fig.~\ref{fig:imaging}(a) shows that part of a cloud initially trapped at the cloud center would be lost to the surface because of the \CP\ force. In contrast, the main part of the figure shows that retaining the magnetic gradient prevents any part of the cloud from being attracted to the surface, and very effectively accelerates the atoms downwards, away from the atomchip.

In Fig.~\ref{fig:imaging}(b) we consider the expansion of a cloud of interacting atoms. The cloud expansion is calculated by solving the time-dependent \GP\ equation for a trapping potential whose frequencies are reduced exponentially as discussed above. This expansion is superimposed on the motion of the center of mass, which is solved classically as shown in the main panel of Fig.~\ref{fig:imaging}(a). By turning the trapping potential off gradually, the cloud expansion is reduced compared to the case where the potential is turned off suddenly. This helps to retain a higher optical density over longer times, allowing the cloud to further separate from the atomchip for detection.  

\section{Conclusion\label{sec:conclusion}}

In this work we have shown that trapping of ultra-cold atoms above surface-grown single-wall carbon nanotubes is feasible with standard~CNT fabrication techniques and electrical parameters. The number of atoms that can be trapped increases with the length of the nanotube and the current passed through it and can reach several thousand for nanotubes longer than~$20\mum$. We have also shown how a practical two-layer CNT-based atomchip may be constructed and how the ultra-cold atoms can be transferred from a conventional magnetic trap to the~CNT trap.

Several types of trap losses were considered. Trap properties were chosen to ensure a Majorana spin-flip loss rate of~$\rm1\,Hz$ at a trap center distance of~$0.85\,\mum$, slow enough to allow experimental confirmation of atom trapping, again using practical elements of the atomchip design. Under these conditions we found that all other trap loss mechanisms are even slower, including tunneling to the atomchip surface, thermal noise-induced spin flips, and heating effects. 

Finally, we show that loading, releasing, and detecting atoms in the~CNT trap suggested here is feasible with simple procedures.

The results of this study should allow the development and testing of CNT-based atomchip traps and the investigation of their properties, including the unique smoothness and electronic characteristics of~CNTs, and perhaps detailed studies of current flow in~CNT contacts. A successful realization of a~CNT atom trap may open the road to numerous improvements of the atomchip including reduced potential corrugation, lower levels of spin flip and decoherence, less destructive absorption near cavity modes thereby facilitating on-chip optical micro-cavities, and, for suspended~CNTs, reduced \CP\ forces and coupling of atoms to mechanical resonators.

\begin{acknowledgments}
We thank the team of the Ben-Gurion University Weiss Family Laboratory for Nanoscale Systems for the fabrication of~CNT-based devices, and J\"urgen Jopp and Roxana Golan of the Ben-Gurion University Ilse Katz Center for Nanoscale Science for assisting with surface measurements. We also thank Ariel Ismach (Weizmann), David Groswasser (BGU), Orel Belcher (BGU), and Yaniv Yarkony (BGU) for assistance with~CNT growth, chip design, and characterization. We are grateful to Marko Burghard (MPI f\"ur Festk\"orperforschung) for discussions on~CNT contacts, to Kim Milton (Oklahoma) for providing the dielectric bilayer \CP\ calculations, and to Carsten Henkel (Potsdam) for critical comments. This work was supported by the European Community ``Atomchip'' Research Training Network and the EC Marie-Curie programme, the American-Israeli Binational Science Foundation, the Israeli Science Foundation, and the Ministry of Immigrant Absorption (Israel).
\end{acknowledgments}


\end{document}